\documentclass[preprint,12pt]{aastex}

\usepackage{amsmath}
\usepackage{txfonts}
\usepackage{cancel}

\begin{document}

\slugcomment{
\sc prepared to be submitted to
\sl the Astrophysical Journal Letters}

\title{Accurate Mass Estimators for NFW Halos}
\author{N.~W.~Evans\altaffilmark{1}, J.~An\altaffilmark{2}, and
A.~J.~Deason\altaffilmark{1}} 
\altaffiltext{1}
{Institute of Astronomy, University of Cambridge,
Madingley Rd., Cambridge~CB3~0HA, the United Kingdom}
\altaffiltext{2}
{National Astronomical Observatories, Chinese Academy of Sciences,
A20 Datun Rd., Chaoyang Dist., Beijing~100012, P.R.China}

\email{
nwe@ast.cam.ac.uk, jinan@nao.cas.cn, ajd75@cam.ac.uk}

\begin{abstract}
  We consider the problem of estimating the virial mass of a dark halo
  from the positions and velocities of a tracer population. Although a
  number of general tools are available, more progress can be made if
  we are able to specify the functional form of the halo potential
  (although not its normalization). Here, we consider the
 Navarro-Frenk-White (NFW) halo and develop two simple estimators. We
 demonstrate their effectiveness against numerical simulations and use
 them to provide new mass estimates of Carina, Fornax, Sculptor, and
 Sextans dSphs.
\end{abstract} 
\keywords{galaxies: halos --- galaxies: kinematics and dynamics
--- dark matter}

\section{Introduction}

There are many instances when we wish to estimate the mass of a dark
halo based on properties of a tracer population. This includes
calculating the masses, of the Milky Way and M31 halos from the
positions and velocities of satellite galaxies, globular clusters and
halo stars; of the dark halos of dwarf spheroidal galaxies from the
projected positions and velocities of constituent stars; of galaxy
clusters from the motions of member galaxies.

There has been much previous work on this problem. Early estimators
were based on the virial theorem, but \citet{BT81} showed that such
estimators are inefficient, biased and, inconsistent, and replaced
them with the projected mass estimator. Their original estimator was
tailored for test particles around a point mass whereas \citet{HTB}
extended it for self-consistent cases. This was modified by
\citet{WEA} to deal with tracer populations whose number density does
not necessarily trace an extended underlying dark matter distribution.
\citet{AE11} on the other hand devised a general theory of mass
estimators under the assumption that the dark matter density law is
known, although the normalization is not.

Here, we start by accepting that numerous studies have shown us
that the dark halos have the universal form,
namely \citep[henceforth NFW]{NFW},
\begin{equation}
\rho(r) 
    \propto \frac1{ r\, (a + r)^2 }
\end{equation}
where $a$ is the scale radius. We wish to find the virial mass
$M_\mathrm{vir}$, which is the mass within the virial radius
$r_\mathrm{vir}$. We deduce the enclosed mass $M(r)$ within the
radius $r$;
\begin{equation}
  M(r) \propto \mu\biggl( \frac{r}{a} \biggr)
\,;\qquad
 \mu(x) \equiv \log(1+x) - \frac{x}{1+x}
\end{equation}
and the normalized mass profile $\tilde M(r)$ follows as
\begin{equation}\label{eq:mprof}
\tilde{M}(r) \equiv \frac{ M(r) }{ M_\mathrm{out} }
             = \frac{ \mu(\tilde{r}) }{ \mu(c) }
\qquad
\text{($0 \le r \le r_\mathrm{out} = r_\mathrm{vir}$)}.
\end{equation}
Here $M_\mathrm{out} = M(r_\mathrm{out})$, and $r_\mathrm{out}$ is the
outermost data point in our sample while $\tilde{r} = r/a =
cr/r_\mathrm{vir}$, and $c = r_\mathrm{vir}/a$ is the concentration
parameter. We have here assumed $r_\mathrm{out} = r_\mathrm{vir}$
and so $M_\mathrm{out}=M_\mathrm{vir}$, which is the case for halo
mass estimation using satellite galaxies.

The problem therefore is: Given an assumed NFW profile together with
the positions and velocities of tracers that extend out to the virial
radius, how can we estimate the virial mass?  The aim of this {\it
  Letter} is to provide a ready-to-use specifically-optimized
estimator for the most popular halo model with tests and applications,
without burdening the reader with a plethora of mathematical
derivations.

\section{Two Estimators}

\citet{AE11} show that if we are willing to assume the functional form
of the halo mass profile $\tilde{M}(r)$ within the given spherical
region of interest, then the total dark halo mass $M_\mathrm{vir}$ in
the same region can be estimated through a weighted average of
kinematic properties of the tracers, adjusted by the boundary term.
The boundary term is related to the external pressure support of the
tracer population, but it is usually taken as small compared to other
uncertainties and so neglected. The mass of the NFW halo is formally
infinite. This means that the profile needs to be truncated at some
radius, which is usually taken to be either $r_\mathrm{out}$ or
$r_\mathrm{vir}$. In this paper, we suppose that the tracers are
well-populated out to the virial radius so that $r_\mathrm{out} =
r_\mathrm{vir}$.

We consider two cases. First, as for the Milky Way satellites, 3-d
distances $r$ to and radial velocities $v_r$ with respect to the halo
center are available for tracers. Strictly speaking, the radial
velocity is with respect to the Sun, but for most practical proposes,
this is almost equivalent to the the radial velocity with respect to
the Galactic Center for distant tracers. Second, as for populations in
most external galaxies, projected distances $R$ and line-of-sight
velocities $v_\ell$ are available for tracers. We consider these
two in turn.

\subsection{True Distances and Radial Velocities}

The general principle underlying all estimators is that the mass is a
weighted average of the positions and velocities of the tracers. For a
given choice of functional form for the dark matter potential,
\citet{AE11} showed that there is an optimum weight function. For the
case of this NFW halo, this gives
\begin{equation}\label{eq:gm}
GM_\mathrm{vir} =
 \biggl\langle \frac{ r w(r) }{ \tilde{M}(r) } v_r^2 \biggr\rangle
-\bcancel{\frac{ 4 \pi r_\mathrm{vir}^4 }{N}
          \nu\sigma_r^2|_{r=r_\mathrm{vir}}},
\end{equation}
where
\begin{equation}
w(r) = 4 - 2 \beta(r)
     - \frac{1}{ \mu(r/a) } \biggl( \frac{r}{a+r} \biggr)^2.
\end{equation}
Here,
$\beta$ is the Binney anisotropy parameter for the spherical system.
\citet{De11} analyzed properties of velocity distributions of
satellites galaxies in numerical simulations of dark halos and found
that typically $\beta \approx 0$. We retain $\beta$ in our formulae,
but in applications we assume $\beta=0$. The last term in equation
(\ref{eq:gm}) is the boundary condition, which is discarded here.
This gives the estimator as
\begin{equation}
\frac{M_\mathrm{vir}}{\mu(c)}
 \simeq \frac1{GN}
        \sum_{i=1}^N \frac{ w(r_i) }{ \mu(r_i/a) }\, r_i v_{r,i}^2
\label{eq:massestone}
\end{equation}
where the index $i$ runs over the $N$ tracers. We refer to this as
the NFW Mass Estimator (NFWME).

It is useful to compare this with a special case of the estimator
derived by \citet{WEA}. They studied dark halos with density profiles
that are scale-free out to the virial radius
\begin{equation}\label{eq:musf}
\tilde{M}(r) = \biggl( \frac{r}{r_\mathrm{vir}} \biggr)^{1-\alpha}.
\end{equation}
For this case, the weight function is a power-law, and the mass
estimator becomes
\begin{equation}\label{eq:ssp}
M_\mathrm{vir} =
 \frac{ (\alpha+\gamma-2\beta)\, r_\mathrm{vir}^{1-\alpha} }{G}
 \left\langle r^\alpha v_r^2 \right\rangle
\end{equation}
where $\gamma$ is the power-law index for the tracer density ($\propto
r^{-\gamma}$). \citet{AE11} have shown that the choice $\gamma = 3$
corresponds to neglecting the boundary term. By contrast, \citet{WEA}
circumvented the effect of the boundary term by explicitly solving for
the power-law solution for the tracers. Here, we recommend the choice
$\gamma \approx 2.5$ for satellite galaxy populations,
but $\gamma \approx 3$ for
stellar populations in dark haloes. Simulations typically find that
the number density of satellite galaxies falls off more slowly than
the stars \citep[e.g.,][]{De11}.  In addition, \citet{WEA} also
argued that $\alpha \approx 0.5$ is a good approximation to the NFW
halo, at least for the range of virial masses and concentrations
appropriate to large galaxies like the Milky Way. With these
particular indices, the mass estimator is in the form:
\begin{equation}
M_\mathrm{vir} \approx
 \frac{ r_\mathrm{vir}^{0.5}\,(0.5 + \gamma - 2\beta) }{GN} 
 \sum_{i=1}^N r_i^{0.5} v_{r,i}^2.
\label{eq:massesttwo}
\end{equation}
We refer to this as the Scale-Free Mass Estimator (SFME). Equations
(\ref{eq:massestone}) and (\ref{eq:massesttwo}) are the fundamental
results of this subsection. They provide simple formulae for the mass
in terms of weighted sums of the positions and velocities. We shall
test these formulae against simulations shortly.

\subsection{Projected Distances and Line-of-Sight Velocities}

In many situations, the radial velocities and true positions of the
tracers with respect to the center of the halo are not direct
observables, but the line-of-sight velocities and projected positions
are. The adjustment of the estimator for this happenstance is covered
in \citet[Sect.~4]{AE11}.

\begin{figure}
\plotone{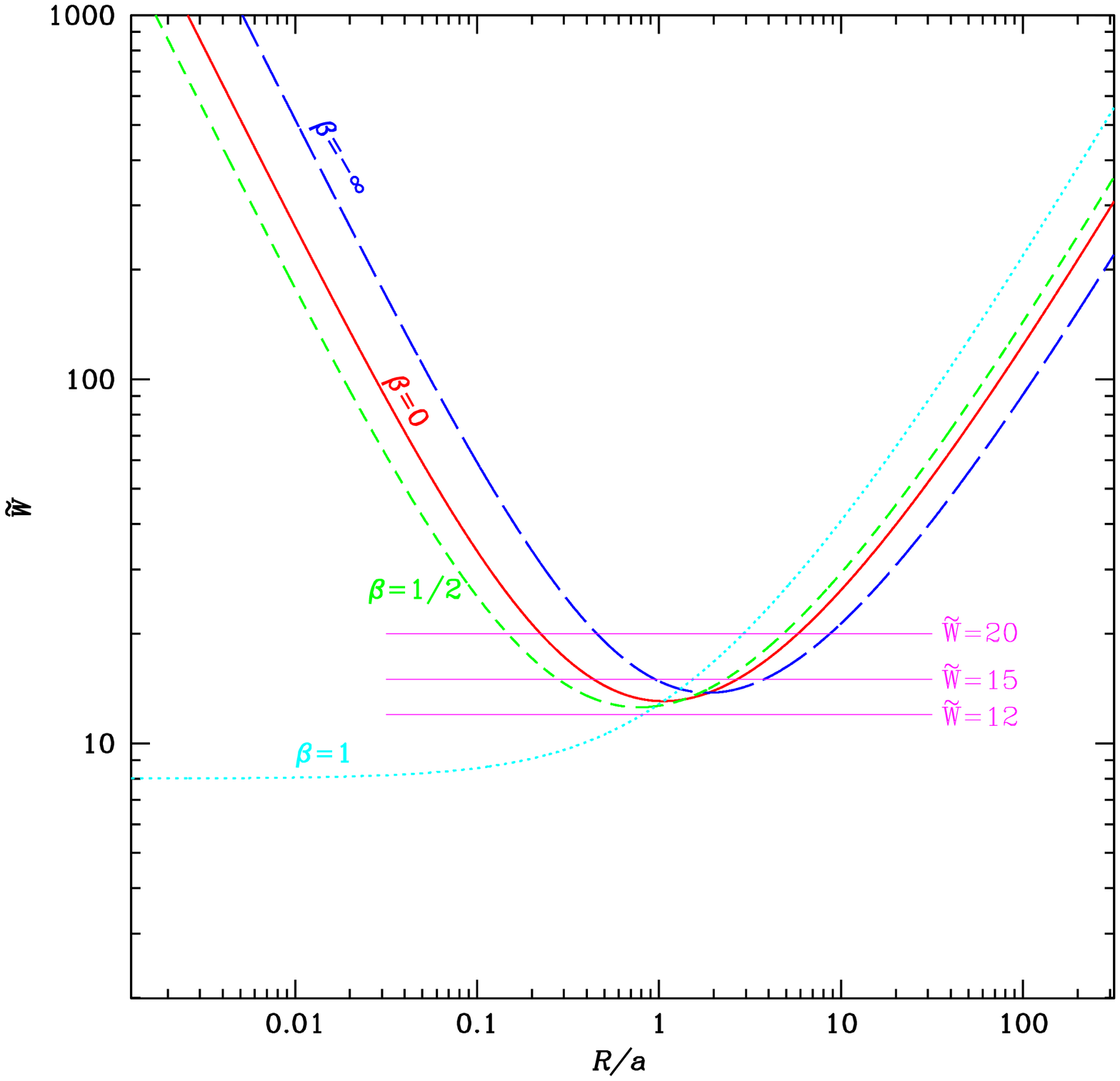}
\caption{\label{fig:w_nfw}
 The invariant normalized weight function $\tilde{W}(\tilde{R})$ for
 the NFW profile as a function of the normalized projected separation
 $\tilde{R} = R/a$. The weight function can be series-expanded at
 $\tilde{R} = 0$ with its leading term behaving as $\sim
 \tilde{R}^{-1}$ unless $\beta = 1$. On the other hand, asymptotically
 toward $\tilde{R} \rightarrow \infty$, we find that
 $\tilde{W}(\tilde{R}) \sim \tilde{R} / \log\tilde{R}$.}
\end{figure}

Here, we would like to find the weight function $W(R)$ of $R$ such
that $\langle v_\ell^2 W(R) \rangle = \langle r w(r) \tilde{M}^{-1}
v_r^2 \rangle$, which is to be substituted into the mass estimator.
\citet{AE11} show that this leads to an integral equation and the
weight function $W(R)$ can be computed, at least numerically. The
invariant (i.e., independent of $c$) normalized weight functions
$\tilde{W}(\tilde{R})\equiv W(R)/[a\mu(c)]$ for the NFW profile for
some constant $\beta$ are shown in Figure~\ref{fig:w_nfw}.

Notice that for much of the range occupied by the tracers and of the
physical range of the anisotropy parameter ($1/2 > \beta > -1$ or the
axis ratio of the velocity ellipsoid no more extreme than
$1:\sqrt{2}$), the weight function in Figure~\ref{fig:w_nfw} is
constant to a good approximation, $\tilde W(\tilde R) \approx 15$, so
that
\begin{equation}
\frac{M_\mathrm{vir}}{\mu(c)}
 \simeq \frac{a}{GN}
        \sum_{i=1}^N \tilde{W}\Bigl( \frac{R_i}{a} \Bigr)\, v_{\ell,i}^2 
\approx \frac{15a}{GN} \sum_{i=1}^N v_{\ell,i}^2.
\label{eq:massestthree}
\end{equation}
Interestingly, this indicates that $M(a) \approx 3 G^{-1} a \langle
v_\ell^2 \rangle$ since $M_\mathrm{vir} / \mu(c) = M(r) / \mu(r/a)$
and $\mu(1) \approx 0.2$. In what follows, we however use the
numerically obtained weight function for the mass estimator, which is
the projected analog of equation (\ref{eq:massestone}).

We will again compare this estimator to the projected analog of the
scale-free estimator. If $\beta$ is a finite constant and the mass
profile is in the scale-free form in equation (\ref{eq:musf}), we find
that \citep[c.f.,][eqs.~26 \& 27]{WEA}
\begin{equation}
M_\mathrm{vir} = \frac{ 4 \Gamma\bigl(\frac{\alpha+5}2\bigr) }
                { \sqrt{\pi} \Gamma\bigl(\frac\alpha2+1\bigr) }
\frac{ (\alpha+\gamma-2\beta)\,r_\mathrm{vir}^{1-\alpha} }
     {\alpha+3-(\alpha+2)\beta}
\frac{ \left\langle R^{\alpha} v_\ell^2 \right\rangle }{G}.
\end{equation}
%
Specializing to the case $\alpha \approx 0.5$ and $\gamma \approx 3$, we find
\begin{equation}
M_\mathrm{vir} \approx
  \frac{ 4 r_\mathrm{vir}^{0.5} }{GN}\,
  \biggl( \frac{ 1 - \frac47\beta }{ 1 - \frac57\beta } \biggr)\
  \sum_{i=1}^N R_i^{0.5} v_{\ell,i}^2.
\label{eq:massestfour}
\end{equation}
This is the projected analog of equation
(\ref{eq:massesttwo}). Equations~(\ref{eq:massestthree}) and
(\ref{eq:massestfour}) are the fundamental results of this subsection.

\section{Applications}

\subsection{Numerical Simulations}

We begin by testing our estimators against simulations. The
Galaxies-Intergalactic Medium Interaction Calculation (GIMIC) suite of
simulations is described in detail in \citet{Ca09}. It consists of a
set of hydrodynamical resimulations of five nearly spherical regions
($\sim 20 h^{-1}\ \mbox{Mpc}$ in radius) extracted from the {\it
Millennium Simulation} \citep{Sp05}. \citet{De11} extracted from the
GIMIC simulations a set of galaxies that resemble the Milky Way. The
sample consists of 431 parent halos and 4,864 associated satellite galaxies.

There are a number of ways in which mock catalogs from simulations
differ from the assumptions used to derive the estimators. For
example, dark halos are not generally spherical, infall continues to
the present day, and the observed satellites are not necessarily
virialized and well-described by an equilibrium distribution. Also,
notions of a constant anisotropy are probably idealized, and in
practice the anisotropy will vary with radius. These systematic errors
are larger than random errors on the measurements of velocities and positions.

\begin{figure*}
\plottwo{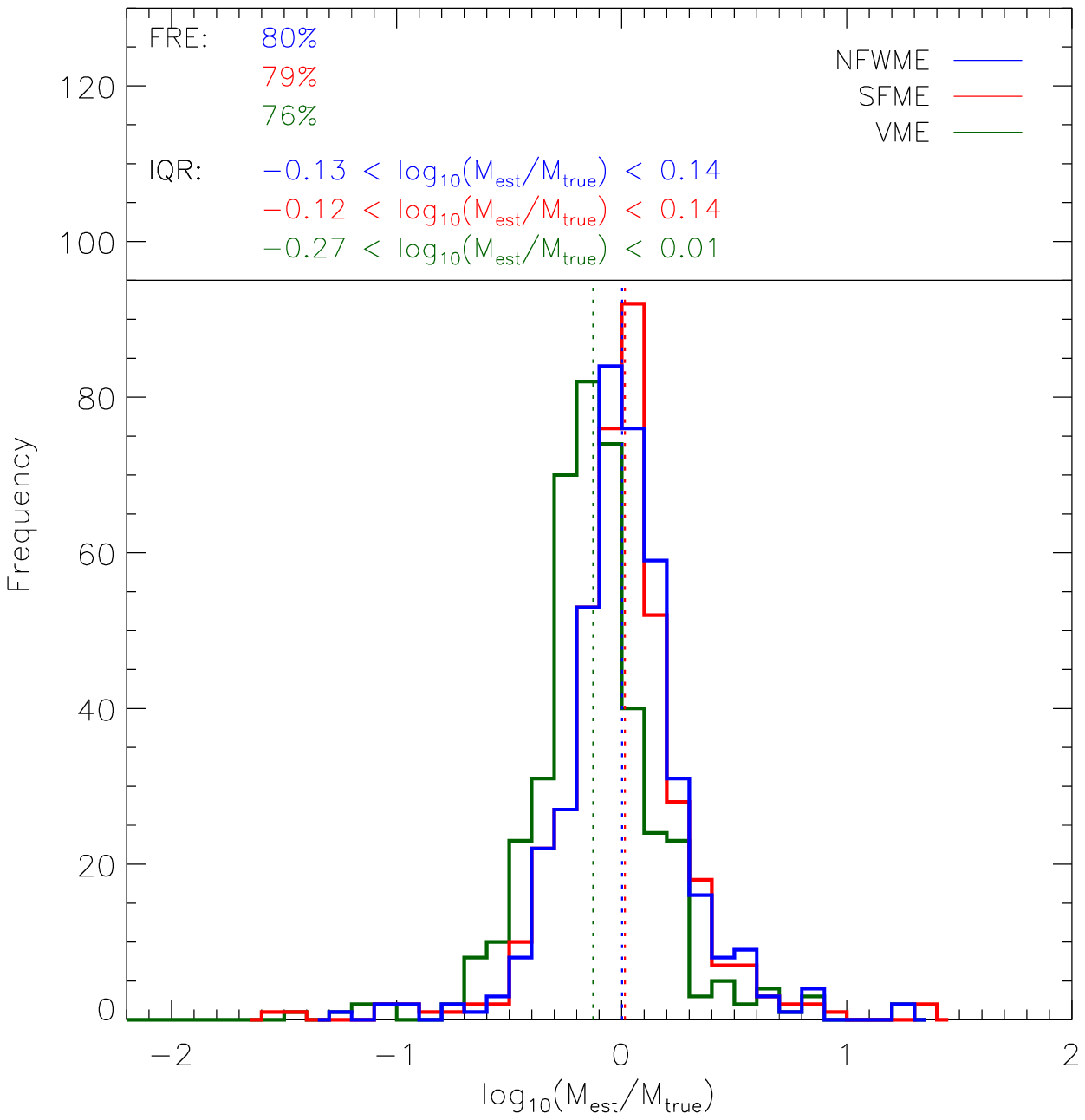}{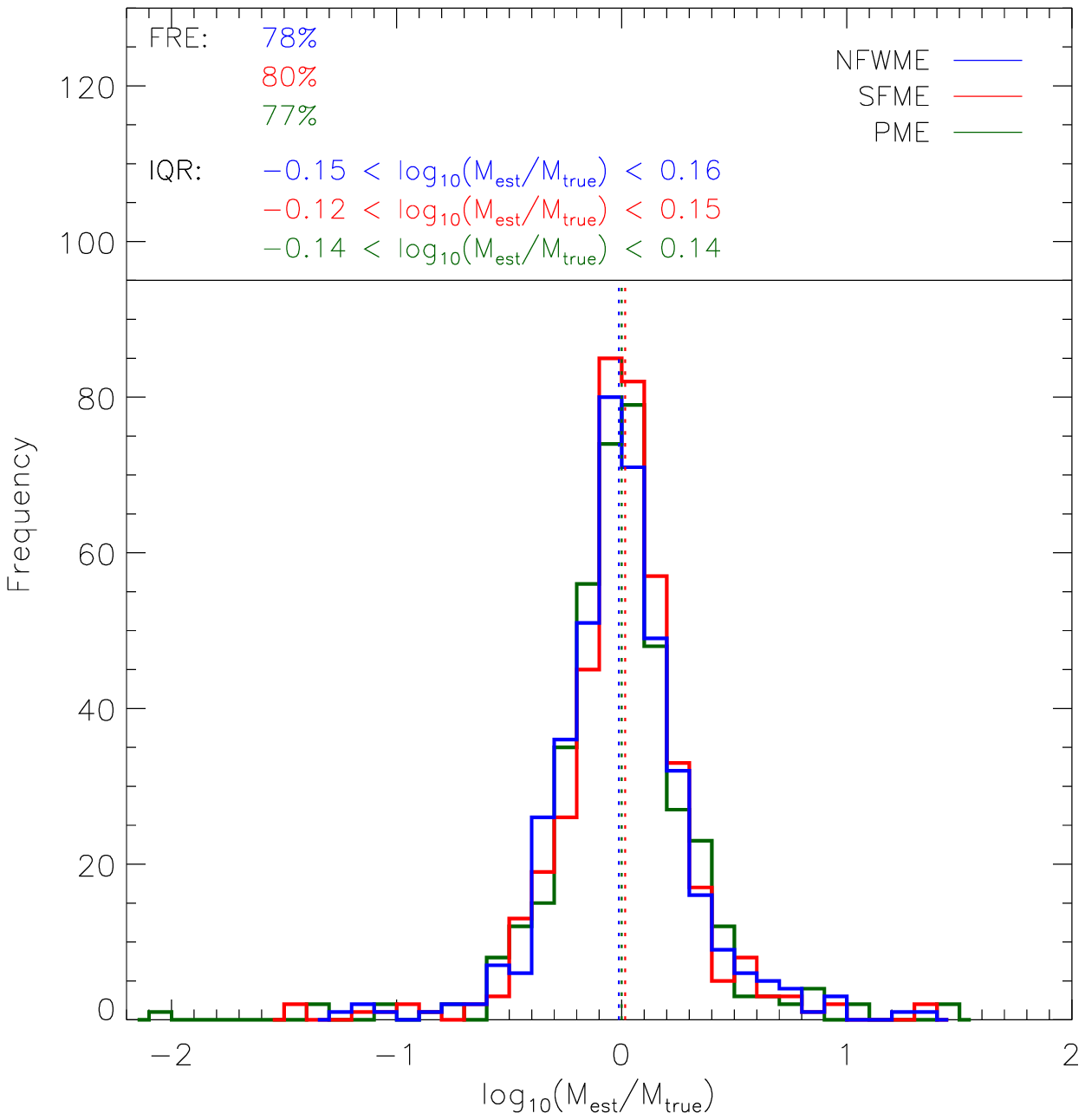}
\caption{\label{fig:alisone} Histograms showing the ratio of the
  estimated mass to the true mass for the two different mass
  estimators. The left panel uses eqs.~(\ref{eq:massestone}) and
  (\ref{eq:massesttwo}) applied to data sets of true distances and
  radial velocities. The right panel uses eqs.~(\ref{eq:massestthree})
  and (\ref{eq:massestfour}) applied to data sets of projected
  distances and los velocities. For the NFWME estimator, a
  concentration $c = 17$ is assumed, which is typical for the haloes
  in the simulations.  FRE and IQR are the fraction of reasonable
  estimates (FRE) and the inter-quartile range (IQR) as defined in the
  text. The means of the distributions are shown by vertical dotted
  lines, from which we see that the virial mass estimator is biased
  and underestimates the true mass (as originally pointed out by
  \citealt{BT81}).}
\end{figure*}

The mass estimators provide an estimate for the total mass within the
radius of the farthest tracer ($\approx r_\mathrm{out}$). We compute
the `true' mass within $r_\mathrm{out} \approx r_\mathrm{vir}$ for
each halo and compare to masses found via our estimators. We use all
satellites, but check that our results are not significantly affected
when only luminous satellites are included. Results are obtained for
the two estimators in this paper -- the NFWME and the SFME -- and also
the virial mass estimator (VME) and the projected mass estimator (PME)
of \citet{BT81}. The left panel of Figure~\ref{fig:alisone} refers to
data sets of radial velocities and true distances, and the right panel
to line-of-sight velocities and projected distances. We quantify the
goodness of any estimator by means of two statistical measures. First,
we define the Fraction of Reasonable Estimates (FRE) as the fraction
of estimates within the factor of two of the true mass
\citep[see also][]{De11}. We also give the Inter Quartile Range (IQR)
of the mass estimates, which gives a good indication of the spread. In
all cases, we see that the NFWME and the SFME outperform the VME and PME.
Interestingly, the performance of the NFWME
is comparable to SFME.
Given this, we recommend the use of equations (\ref{eq:massesttwo})
and (\ref{eq:massestfour}) for practical applications.

\subsection{Dwarf Spheroidal Masses}

We now turn to an astrophysical problem. The recent years have seen
programs to garner the radial velocities of giant stars in the nearby
dwarf spheroidal galaxies (dSphs). For the brighter of the dSphs,
radial velocity surveys have provided data sets of projected positions
and line-of-sight velocities for thousands of stars \citep[see
e.g.,][] {Kl02,Wi04,Wa09a}. This has driven new theoretical ideas,
such as the claim by \citet{St08} that all the dSphs shared a common
mass scale of $\sim 10^7\ \mbox{M$_\sun$}$ within 300~pc. The data has
also driven the study of new techniques for mass estimation and
modeling \citep{Wo10,Am10}.

We use the velocities and positions of individual stars observed in
four classical dSphs -- Carina, Fornax, Sculptor, and Sextans --
presented by \citet{Wa09a}. We select only those stars for which the
probability of membership is $>0.9$, and additionally assume that the
isotropy ($\beta = 0$). The results for the mass within 300 and 600~pc
are listed in Table \ref{tab:res}. There are a couple of points worth
noting. First, the masses do indeed support \citet{St08}'s notion of a
common mass scale within 300~pc, although this is less surprising as
the velocity dispersions of these four dSphs are similar. Second, the
PME agrees well with the two new estimators for the enclosed mass
within 300~pc. However, it begins to diverge from the SFME and NFWME at
larger radii.

Although the results in Table \ref{tab:res} are comparable to mass
estimates obtained through Jeans \citep{Wa09b} and distribution
function \citep{Am10} modeling, the amount of effort involved is very
much less. All that is required is a weighted average of positions and
velocities as opposed to solving differential, or integro-differential
equations.

\section{Conclusions}

We have provided a new and accurate way of estimating the masses of
NFW halos from the positions and radial velocities of tracer
populations. This work follows up the theoretical paper of \citet{AE11}
and supplies a ready-to-go mass estimator for the most common
application of all.

We have supposed that the reader has a data set of $N$ objects with
true positions $r_i$ and radial velocities $v_{r,i}$, or with
projected positions $R_i$ and line-of-sight velocities
$v_{\ell,i}$. In terms of overall simplicity and flexibility, we
recommend using the isotropic limit ($\beta = 0$) with an $r^{-3}$
tracer number density fall-off ($\gamma = 3$)
%
\begin{equation}\begin{split}
M_\mathrm{out} &\simeq
 \frac{3.5 r_\mathrm{out}^{0.5} }{GN} 
 \sum_{i=1}^N r_i^{0.5} v_{r,i}^2
\\
M_\mathrm{out} &\approx
 \frac{  4 r_\mathrm{out}^{0.5} }{GN}
 \sum_{i=1}^N R_i^{0.5} v_{\ell,i}^2
\end{split}\end{equation}
%
to compute the enclosed mass within the sphere of the radius
$r_\mathrm{out}$. Here, $r_\mathrm{out}$ is the location of the
outermost data point. In the projected case, only $R_\mathrm{out}$ is
known, but on statistical grounds, we have that $r_\mathrm{out} = \pi
R_\mathrm{out} / 4$. Obviously, no information can be inferred on the
mass distribution exterior to the outermost data point.  

\acknowledgments

We thank the GIMIC simulators for access to numerical data.  JA thanks
for hospitality during his visits to IoA (Cambridge).  He is supported
by the Chinese Academy of Sciences (CAS) Fellowships for Young
International Scientists, Grant No.:2009Y2AJ7, and the National
Natural Science Foundation of China (NSFC) Research Fund for
International Young Scientists.  AJD is supported by a studentship
from the Science and Technology Facilities Council (STFC) of the
United Kingdom.

\begin{deluxetable}{lcccc}
\tablewidth{0pt}
\tablecaption{Dwarf Spheroidal Masses}
\tablehead{\multicolumn{2}{c}{dSph} & PME  & SFME & NFWME \\
 && \multicolumn{3}{c}{($\times\ 10^7\ \mbox{M$_\sun$}$)} }
\startdata
 Carina   & ($R < 300\ \mbox{pc}$) & 1.7 & 1.7 & 1.7 \\
          & ($R < 600\ \mbox{pc}$) & 2.2 & 2.6 & 2.7 \\
 Fornax   & ($R < 300\ \mbox{pc}$) & 2.9 & 2.8 & 2.9 \\
          & ($R < 600\ \mbox{pc}$) & 5.1 & 5.2 & 5.2 \\
 Sculptor & ($R < 300\ \mbox{pc}$) & 1.8 & 1.8 & 1.8 \\
          & ($R < 600\ \mbox{pc}$) & 2.5 & 3.1 & 3.0 \\
 Sextans  & ($R < 300\ \mbox{pc}$) & 1.3 & 1.3 & 1.3 \\
          & ($R < 600\ \mbox{pc}$) & 2.0 & 2.2 & 2.2
\enddata
\tablecomments{All the error are $\pm 0.1\times10^7 \mbox{M$_\sun$}$, taking
  into account the random measurement uncertainties only. For the
  NFWME, a concentration $c=30$ (typical for dSphs) has been assumed.}
\label{tab:res}
\end{deluxetable}

\end{document}